\documentclass[aps,prb,onecolumn]{revtex4}
\usepackage{tabularx,graphicx}
\usepackage{amsmath, amsthm, amssymb} 
\usepackage{bbm}
\renewcommand{\r}{{\bf r}}
\newcommand{\q}{{\bf q}}
\renewcommand{\d}{{\bf d}}

\renewcommand{\t}{{\bf t}}
\newcommand{\K}{{\bf K}}

\renewcommand{\v}{{\bf v}}

\renewcommand{\k}{{\bf k}}
\renewcommand{\b}{{\bf b}}
\renewcommand{\a}{{\bf a}}
\newcommand{\G}{{\bf G}}  
\begin{document}

\title{Dirac-point engineering and topological phase transitions \\
in honeycomb optical lattices}
\author{B. Wunsch$^{1,2}$, F. Guinea$^{2}$, F. Sols$^{1}$}
\affiliation{$^1$ Departamento de F\'isica de Materiales, Universidad Complutense de Madrid, E-28040
  Madrid, Spain.\\$^2$ Instituto de Ciencia de Materiales de Madrid, CSIC, Cantoblanco,
  E-28049 Madrid, Spain.}
\date{\today}
\email{bwunsch@fis.ucm.es}
\begin{abstract}
  We study the electronic structure and the phase diagram of non-interacting
  fermions confined to hexagonal optical lattices.  In the first part, we
  compare the properties of Dirac points arising in the eigenspectrum of
  either honeycomb or triangular lattices. Numerical results are complemented
  by analytical equations for weak and strong confinements. In the second part
  we discuss the phase diagram and the evolution of Dirac points in honeycomb
  lattices applying a tight-binding description with arbitrary nearest-neighbor hoppings. With increasing asymmetry between the hoppings the Dirac
  points approach each other. At a critical asymmetry the Dirac points merge
  to open an energy gap, thus changing the topology of the eigenspectrum. We
  analyze the trajectory of the Dirac points and study the density of states
  in the different phases.  Manifestations of the phase transition in the
  temperature dependence of the specific heat and in the structure factor are
  discussed.
\end{abstract}

\maketitle
\section{Introduction}
The isolation of a single layer of graphene\cite{Nov04}, which is a
two-dimensional honeycomb lattice of carbon atoms, has attracted considerable
attention, both for its basic interest and because it may pave the way
to carbon-based electronics. The low energy electronic properties are
described by the massless Dirac equation, which causes many anomalies with
respect to semiconductor physics\cite{rmp,GN07,BeenakkerRev}.  Close to
half-filling, the band structure near the Fermi energy is given by two
degenerate Dirac cones, and the centers of these cones (called Dirac points)
are located at the two distinct K-points of the first Brillouin zone. A
first experimental validation of the relativistic band structure was the
measurement of the anomalous sequencing of the Quantum Hall
plateaus\cite{Zhang05,Novoselov05}.

In graphene there is little room to tune system parameters such as the
interaction strength or the hopping amplitudes, which prevents a systematic
study of its phase diagram. In contrast for cold fermionic atoms confined in
honeycomb optical lattices\cite{Grynberg93} the parameters are highly tunable
and many of the theoretically predicted phases might be realized there. Recent
theoretical works on honeycomb lattices have studied a topological phase
transition in the single particle spectrum which is due either to asymmetric
hopping energies\cite{Duan07}, or to a Kekul\'{e} distortion of the
hoppings\cite{Mudry07,Herbut07}, or to the application of an ac electric
field\cite {Zhang08}. Further work on honeycomb optical lattices has dealt
with the realization of the anomalous Hall and the Spin Hall
effects\cite{Dudarev04}, the Wigner crystallization of $\pi
$-bands\cite{Wu07,Wu07a}, or with the Quantum Hall effect without Landau
levels\cite{Wu08b,Shao08,Haldane88}.  Finally, we note recent theoretical work
on the appearance and manifestation of Dirac cones in the band structure of
triangular photonic crystals\cite{Raghu05,Raghu06,Beenakker08}.

There are two main purposes of this work. The first one is to compare the Dirac
points present in the lower lying bands of triangular and honeycomb lattices.
Therefore we first discuss in section II the potential landscape of these
lattices.  Then, in section III we discuss the Dirac cones appearing in both
lattice types and derive analytical results within the nearly-free-particle
and tight-binding limits.  The second main goal of this work is to analyze in
detail the topological phase transition in honeycomb lattices with asymmetric
hopping energies\cite{Duan07}. In section IV, the evolution of the Dirac
points and the resulting phase diagram for the honeycomb lattice as a function
of the relative tunnel couplings are studied within the tight-binding limit .
The density of states for the different phases is derived in section V.
Experimental signatures of the phase transition such as the structure factor
and the temperature dependence of the specific heat are explained in section
VI. Finally, we discuss the topological structure of the phase transition in
section VII. Some technical derivations are left to the Appendices.

\section{Potential landscapes}
\begin{figure}
  \begin{center}
    \includegraphics*[angle=0,width=0.8\linewidth]{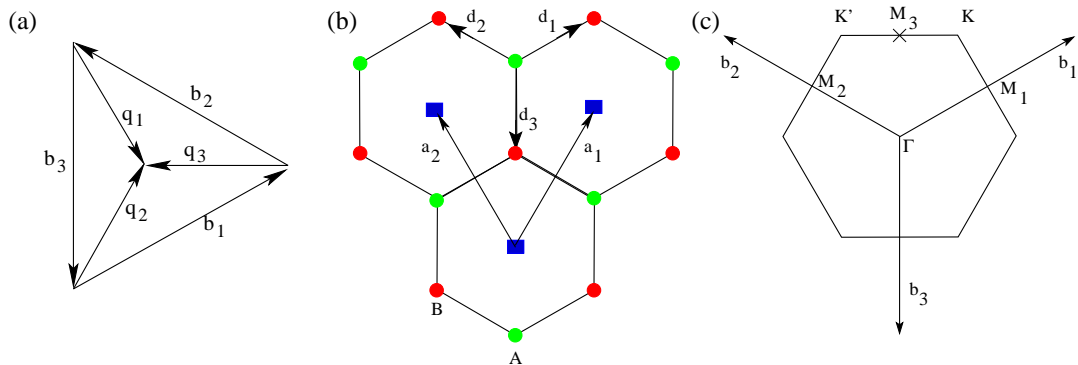} 
    \caption{(a) Three traveling lasers with wavevectors $\q_i$ interfere
      and build up the hexagonal lattice with reciprocal unit vectors
      $\b_1,\b_2$, and $\b_3=-(\b_1+\b_2)$. (b) Unit cell in real space spanned by
      the unit vectors $\a_1,\a_2$. For triangular lattices the minima are
      located at
      the center of the hexagons, while for symmetric honeycomb lattices
      minima lie at the $A$ and $B$ sites. (c) Unit cell in reciprocal
      space (first Brillouin zone) indicating the position of the symmetry points
      $\Gamma,{\rm M_1,M_2,M_3}$ and the K-points.}
\label{fig:Unitcells}
\end{center}
\end{figure}
We consider the optical lattice that was experimentally realized by Grynberg
{\it et al.} \cite{Grynberg93}. It is created by three interfering traveling
laser beams. The corresponding wave vectors are in the $x-y$ plane and form
equal angles between them, as illustrated in Fig.~\ref{fig:Unitcells}a.  For
simplicity we assume the polarization of the beams to be perpendicular to the
plane, although the confining potential is independent of the polarization
for large enough detuning\cite{Dudarev04}.

The amplitude of the total electric field (polarized in the $z$-direction) and the light intensity can then be written
\begin{eqnarray}
E(\r)&=&\sum_j E_j \exp(i \q_j\cdot \r) \\
I(\r)&=&|E(\r)|^2=E_1^2+E_2^2+E_3^2+2 E_1 E_2 \cos(\b_3\cdot \r)+2 E_2 E_3
\cos(\b_1\cdot \r)+2 E_3 E_1 \cos(\b_2\cdot \r)\label{eq:I}\;.
\end{eqnarray}
As illustrated in Fig.~\ref{fig:Unitcells}a $\b_1=\q_2-\q_3$ plus cyclic permutations. 
We note that the system is essentially insensitive to phase drifts between the lasers,
since these just shift the lattice without changing the potential landscape. We can
therefore assume $E_i>0$ without loss of generality.

The optical confining potential $V(\r)$ is proportional to the intensity of
the laser field and the proportionality factor is positive (negative) if the
laser frequencies are blue (red) detuned with respect to the
transition frequency of the atoms. Up to a constant the confinement is given by
\begin{eqnarray}
V(\r)&=&\sum_j V_j \cos(\b_j\cdot \r), \label{eq:Pot}
\end{eqnarray}
with $V_1\propto E_2 E_3$ plus cyclic permutations and $V_j>0$ ($V_j<0$) for
blue (red) detuning.

The potential $V(\r)$ has an underlying triangular Bravais lattice, whose
lattice spacing $a$ is determined by the wavelength of the traveling
lasers $a=4 \pi/3 q=2 \lambda/3$.  The unit vectors for real and reciprocal
lattices are given by:
\begin{eqnarray}
\a_1&=&\frac{a}{2}\left(\begin{array}{c}1\\\sqrt{3} \end{array}\right)\,;\quad
\a_2=\frac{a}{2}\left(\begin{array}{c}-1\\\sqrt{3}
  \end{array}\right)\,;\quad\b_1=\frac{2\pi}{\sqrt{3} a}\left(\begin{array}{c}\sqrt{3}\\1 \end{array}\right)\,;\quad
\b_2=\frac{2\pi}{\sqrt{3} a}\left(\begin{array}{c}-\sqrt{3}\\1
  \end{array}\right)\quad \label{eq:bvector2}
\end{eqnarray}
with $\b_i\cdot \a_j=2\pi \delta_{ij}$. The unit cells of both reciprocal and real
space can be chosen of honeycomb form as shown in Figs~\ref{fig:Unitcells}b
and c.

In the case of red detuning, the potential minima lie at the centers of the
hexagons thus forming a triangular lattice for all values of $V_j<0$.  If the
lasers are blue detuned and their intensities are equal, then the minima of
the potential $V(\r)$ are located at the vertices of the honeycombs while
there is a maximum of the potential at the center of each honeycomb, see
Fig.~\ref{fig:Unitcells}b. If the intensities $V_j$ become different (i.e. the
electric field amplitudes $E_j$ begin to be different), the two minima per
unit cell approach each other until they merge for strong asymmetries, namely,
when $|E_1-E_2|>E_3$ or $E_3>E_1+E_2$.  Thus the potential
$V(\r)$~(\ref{eq:Pot}) has two minima per unit cell for
$|E_1-E_2|<E_3<E_1+E_2$.  It is evident from Eq.~(\ref{eq:Pot}) that the
potential $V(\r)$ has inversion symmetry, which is crucial for the stability
of the Dirac points\cite{Guinea07}. The position of the minima are given by
$\r=\pm \r_{\rm min}$ with:
\begin{eqnarray}
\r_{\rm min}=\left(\begin{array}{c} \frac{a}{2 \pi}\;\text{sgn}(E_2-E_1) \arccos\left(\frac{E_1+E_2}{E_3}\sqrt{\frac{E_3^2-(E_2-E_1)^2}{4E_1 E_2}}\right)\\\frac{a\sqrt{3}}{2 \pi}\;\arccos\left(-\sqrt{\frac{E_3^2-(E_2-E_1)^2}{4E_1
          E_2}}\right)\end{array}\right)\label{eq:rmin}
\end{eqnarray}  
We will show in section IV that the energy dispersion $E_\k$ calculated within the
tight-binding approximation has the same structure as
the potential $V(\r)$. The analog of the motion of the potential minima for the case of
the energy spectrum is the evolution of the Dirac points (\ref{eq:DiracP}), which will finally
merge, causing a topological phase transition from a semimetal to an insulator. 

\section{Symmetric lattices}
\begin{figure}
  \begin{center}
    \includegraphics*[angle=0,width=0.9\linewidth]{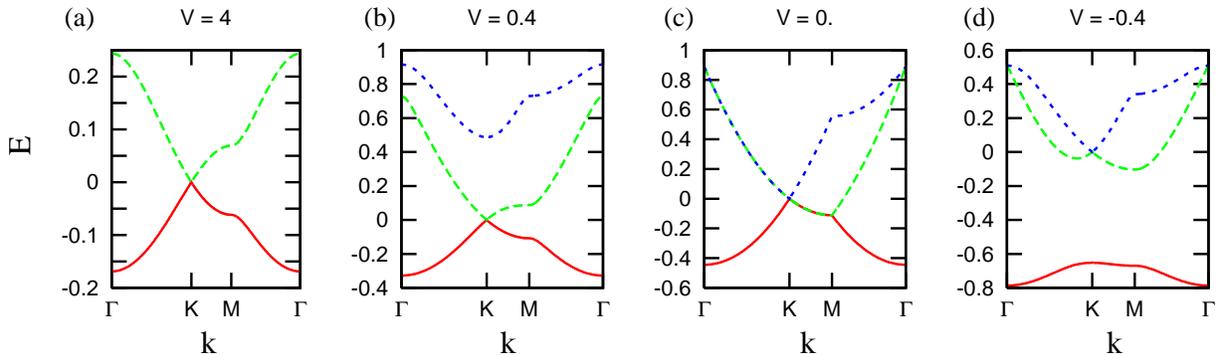} 
    \caption{Band structure for the symmetric lattice ($V_1=V_2=V_3=V$) along
      characteristic lines of the first Brillouin zone for various values of
      $V$. Symmetry points are explained in Fig.~\ref{fig:Unitcells}c.
      The energy of Dirac point is set to 0 and the unit of energy is $(\hbar^2/2m)(4\pi^2/a^2)$. }
\label{fig:Band}
\end{center}
\end{figure}

Because of periodic translational invariance, the crystal momentum is
conserved and the eigenspectrum of a particle moving in the potential $V(\r)$
of Eq.~(\ref{eq:Pot}) is most conveniently calculated in Fourier space.  The
eigenfunctions are thus written as $\psi_\q(\r)=\sum_{m,n} c_{\q-\G_{mn}}
\exp(i(\q-\G_{mn})\cdot \r)$, where $\q$ is restricted to be in the first
Brillouin zone, $\G_{mn}=m \b_1+n \b_2$ denotes a general reciprocal lattice
vector, and the potential is written as $V(\r)=\sum_{mn} V_{mn} \exp(i
\G_{mn}\cdot \r)$ where $V_{mn}=\frac{1}{2}\left[V_3
  (\delta_{m,1}\delta_{n,1}+\delta_{m,-1}\delta_{n,-1})+V_1 \delta_{n,0}
  (\delta_{m,1}+\delta_{m,-1})+V_2 \delta_{m,0}
  (\delta_{n,1}+\delta_{n,-1})\right]$ are the Fourier components of the
potential. Due to inversion symmetry of the lattice $V_{mn}=V_{-m,-n}$.

For each $\q$ the Fourier components $c_{\q-\G_{mn}}$ form a vector $\v_\q$,
where each entry is labeled by a specific pair of integers
$(\v_\q)_{mn}=c_{\q-\G_{mn}}$.  The Schr\"odinger equation then reduces to a set
of linear equations for each $\v_\q$ given by:
\begin{eqnarray}
{\bf M}_\q \v_\q&=&E_\q \v_\q \label{Eq:Eigenspectrum}\\
({\bf M}_\q)_{(m_1,n_1),(m_2,n_2)}&=&\delta_{m_1,m_2}\delta_{n_1,n_2}
E^0_{\q-\G_{m_1,n_1}}+V_{m_1-m_2, n_1-n_2}\notag
\end{eqnarray}
where $E^0_{\q}=\hbar^2 q^2/2m$ denotes the kinetic energy.

Figure~\ref{fig:Band} shows the band structure for symmetric potentials
$V_1=V_2=V_3=V$ for various confinement strengths $V$. Most importantly, we
note that, both for triangular and honeycomb lattices, some pairs of bands
touch at the K-points and in the vicinity of those touching points the band
structure is given by cones, which are called Dirac cones because of the
similarity to the relativistic energy dispersion.

There are however important differences between both cases. For the honeycomb
lattice the first pair of Dirac points (located at $\pm\K$) arises due to the
touching of the two lowest bands.  Furthermore, the Fermi surface in the
relevant energy interval is exclusively determined by these Dirac cones,
leading in particular to a vanishing density of states at complete filling of
the lowest band.

By contrast, for a triangular lattice the first pair of Dirac points is caused
by the touching between the second and third band. Since these bands are
also degenerate at the $\Gamma$ point, the Dirac points are not isolated but
resonate with a continuous band that dominates the density of states in the
energy regime of the Dirac cones.  Furthermore, the filling factor (number of
atoms per unit cell) at which the Fermi energy coincides with the Dirac point
is not an integer, also in contrast to the honeycomb lattice.

The formation of the Dirac cone can be derived within the nearly-free-particle
approximation, where the potential is treated perturbatively as shown in
Appendix~\ref{app:NFE}. The distinction between honeycomb and triangular
lattices is however better appreciated in the tight-binding regime, where the
Hamiltonian is written in terms of Wannier states, each localized at a
potential minimum as shown in Appendix~\ref{app:TB}.

In the following we concentrate on the honeycomb lattice within the tight-binding approximation.

\section{Dirac points in asymmetric honeycomb lattices}
\begin{figure}
  \begin{center}
    \includegraphics*[angle=0,width=0.8\linewidth]{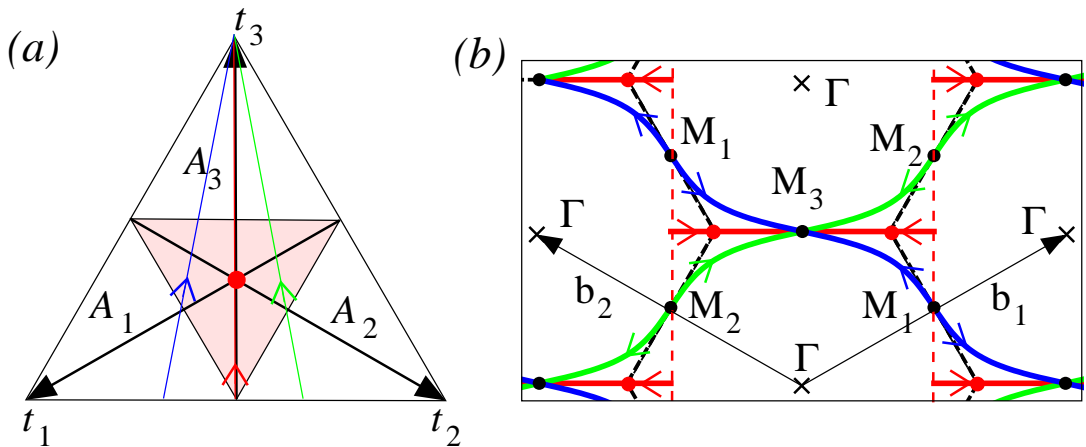} 
    \caption{(a) Phase diagram of the asymmetric honeycomb lattice within
      tight-binding description.  Black lines with arrows are coordinate axes
      for the tunneling amplitudes.  Pink central region denotes semimetal
      with two distinct Dirac points. White regions labeled by $A_1,A_2,A_3$
      denote insulating phases, in which no Dirac points are present. In the
      insulating phase there is a single band minimum in the first Brillouin
      zone located at M$_1$ for $A_1$ and so forth.  (b) Motion of Dirac points
      as a function of $t_3$ for $t_1/t_2=1$ (red line in a and b)
      $t_1/t_2=3/2$ (blue) $t_1/t_2=2/3$ (green). Arrows in (a) and (b) indicate
      direction of motion of Dirac points for increasing $t_3$. The symmetric
      case $t_1=t_2=t_3$ (like in graphene) is indicated by red circles.}
\label{fig:Phasediag}
\end{center}
\end{figure}

We now discuss the eigenspectrum of a honeycomb lattice with asymmetric
hoppings, described by the following tight-binding Hamiltonian:
\begin{eqnarray}
H_\k &=&- \left(a_\k^\dag,b_\k^\dag\right)\left(\begin{array}{c c}0 & \xi_\k\\
  \xi_\k^* & 0\end{array}\right) \left(\begin{array}{c} a_\k\\
  b_\k\end{array}\right)\,;\quad  \xi_\k=\sum_j t_j \exp(-i \k \cdot \d_j) \label{eq:Hasym}
\end{eqnarray}
where $a_\k,b_\k$ destroy Bloch waves on the two different sublattices and
$t_j>0$ denote nearest-neighbor hopping energies. For symmetric hoppings the
standard Hamiltonian~(\ref{eq:Hsym}) is rediscovered.

A possible realization of this Hamiltonian is a deep optical potential
$V(\r)$~(\ref{eq:Pot}) with different values of $V_j$. Due to the inversion
symmetry of the confinement potential~(\ref{eq:Pot}) the onsite energies at
the two sublattices are equal and thus can be set to zero in
Eq.~(\ref{eq:Hasym}). For asymmetric laser intensities the potentials barriers
separating a given potential minimum from the three nearest minima are
different. In particular, the corresponding distances between neighboring
minima are also different, as follows from Eq.~(\ref{eq:rmin}). Both effects
will lead to asymmetric nearest-neighbor hoppings. We note that a pure shift
in the position of the potential minima without a change in the magnitude of
the tunnel hopping energies only gives rise to a phase factor in
$\xi_\k$~(\ref{eq:Hasym}) and thus does not affect the energy spectrum
$E_\k=\pm |\xi_k|$.

The energy spectrum of the Hamiltonian~(\ref{eq:Hasym}) is symmetric around
zero energy and is given by $E_\k=\pm |\xi_k|$. We note the close analogy
between the energy amplitude $\xi_\k$ and the total electric field $E(\r)$.
This causes a close similarity between $E_\k^2=|\xi_\k|^2$ and $V(\r)\propto
E(\r)^2$. As discussed above for the symmetric case $t_1=t_2=t_3$, valence and
conduction band (defined by negative and positive energies, respectively)
touch at the two K-points, where they form Dirac cones. Introducing an
asymmetry between the tunneling amplitudes the two Dirac points move away from
the K-points and approach each other. At a critical asymmetry the Dirac points
merge at one of the three inequivalent M-points, and by increasing the
asymmetry further a band gap opens.

Figure~\ref{fig:Phasediag} illustrates the phase diagram and the displacement
of the Dirac points as a function of the tunnel amplitudes. The central pink
region in Fig.~\ref{fig:Phasediag}a embraces the parameter regime of the
semimetallic phase, where the asymmetry is small enough for the two
inequivalent Dirac points to exist. It is defined by the condition
$|t_{1}-t_{2}|<t_{3}<t_{1}+t_{2}$, which remains invariant under the
permutation of the three subindices. The Dirac points are located at $\k=\pm
\k_{D}$, with:
\begin{eqnarray}
\k_D=\left(\begin{array}{c} \frac{2}{a} \arccos\left(-\sqrt{\frac{t_3^2-(t_2-t_1)^2}{4t_1
          t_2}}\right)\\\frac{2}{a \sqrt{3}}\;\text{sgn}(t_1-t_2)
    \arccos\left(\frac{t_1+t_2}{t_3}\sqrt{\frac{t_3^2-(t_2-t_1)^2}{4t_1
          t_2}}\right)\end{array}\right)
\label{eq:DiracP}
\end{eqnarray}
If necessary the wavevector $\k_D$ as defined above is supposed to be folded
back to the first Brillouin zone by adding a uniquely defined reciprocal
lattice vector.

Figure~\ref{fig:Phasediag}b shows the displacement of the Dirac points
following three different lines of the phase diagram, each defined by a fixed
ratio $ t_{1}/t_{2}$ while spanned by varying $t_{3}$. We focus first on the
red line, which corresponds to the partially symmetric case $t_{1}=t_{2}$
previously studied in Ref.~[\onlinecite{Duan07}]. For $t_{3}=0$ valence and
conduction band touch along the extended dashed line. Increasing $t_3$ causes
the two inequivalent Dirac cones to approach each other along horizontal lines
and for symmetric hoppings $t_1=t_2=t_3$ (center of triangle in Fig.
\ref{fig:Phasediag}a) they are located at the K-points. Finally the Dirac
cones merge at the symmetry point M$_{3}=(\b_{1}+\b_{2})/2$ when
$t_{3}=t_{1}+t_{2}$.

In the generic case of $t_{1}\neq t_{2}\neq t_{3}$ one finds the motion shown
by the blue and green lines. We now discuss the motion for increasing $t_{3}$.
For $t_{3}<|t_{1}-t_{2}|$ the system is an insulator and there is a single
band minimum located at M$_{1}$, if $t_{1}>t_{2}$ (blue line in region $A_{1}$
of Fig.~\ref{fig:Phasediag}a), or M$_{2}$, if $t_{2}>t_{1}$ (green line in
region $A_{2}$ of the same figure). While in regions $A_{1}$ or $A_{2}$, the
minimum stays pinned at the fixed points M$_{1}$ or M$_{2}$ and motion along
the blue or green line only causes the corresponding gap to decrease until it
becomes zero when the central pink triangle is touched (phase transition). For
$|t_{1}-t_{2}|<t_{3}<t_{1}+t_{2}$, in both cases (blue and green) the system
enters the central triangle of Fig.~\ref{fig:Phasediag}a defining the
semimetallic phase, where the band minimum at M$_{1}$ or M$_{2}$ separates
into two Dirac points. The position of the Dirac points are related to each
other by the inversion symmetry around the symmetry points $\Gamma $ and
M$_{1}$, M$_{2}$, M$_{3}$. Finally at $t_{3}=t_{1}+t_{2}$, where the lines leave the
pink triangle in order to enter region $A_{3}$, the Dirac points merge at
M$_{3}$ and the system undergoes a phase transition from a semimetal to an
insulator. Further increase of $t_{3}$ only causes the gap at M$_{3}$ to
increase.

\section{Density of states}
\begin{figure}
  \begin{center}
    \includegraphics*[angle=0,width=0.8\linewidth]{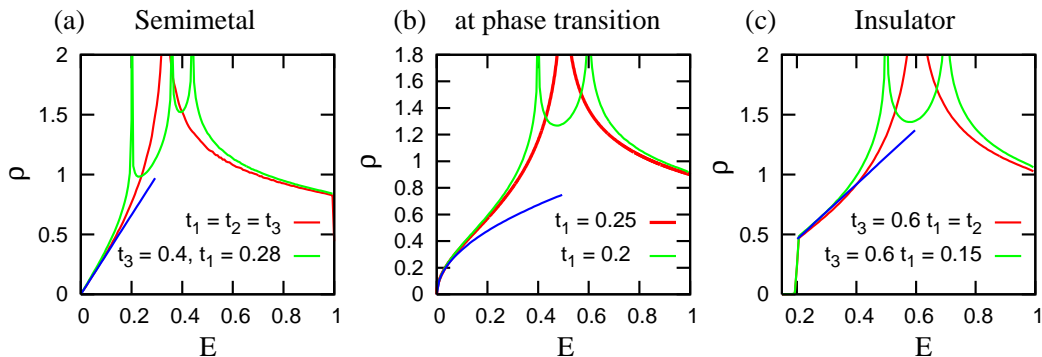} 
    \caption{Density of states per unit cell for different hoppings normalized
      to $t_1+t_2+t_3=1$. (a) Linear energy dependence at low energy within
      semimetallic phase. (b) Square root energy dependence at phase transition
      $t_3=0.5$. (c) Insulating phase. Blue curves are analytical
      approximations to red curves given in main text.}
\label{fig:Dos}
\end{center}
\end{figure}
We now discuss the density of states, which is defined by $\rho(E)=A^{-1} \sum_\k
\delta(E-E_\k)$, where $A=N A_u$ is the area, $N$ the number of unit cells and
$A_u=a^2 \sqrt{3}/2$ the size of the unit cell.  Characteristic energies are
given by the eigenenergies at the symmetry points $E_{\rm M_3}=|t_3-t_1-t_2|$,
$E_{\rm M_1}=|t_1-t_2-t_3|$,$E_{\rm M_2}=|t_2-t_1-t_3|$ and $E_{\Gamma}=t_1+t_2+t_3$.
$E_{\Gamma}$ is the energy maximum and thus sets the bandwidth. The density
has a step at $E=\pm E_\Gamma$.

Without loss of generality we focus on the case $t_3>t_1,t_2$, so that the
energy always has a saddle point at the symmetry points ${\rm M_2,M_3}$, which
results in logarithmic van Hove singularities at $E_{\rm M_2},E_{\rm M_3}$.
Whether M$_3$ is a saddle point or a minimum depends on the tunnel amplitudes.
For $t_3<t_1+t_2$ the energy has a saddle point at M$_3$ and there are two
Dirac points in the first Brillouin zone. For symmetric hoppings $t_3=t_1=t_2$
the band structure close to zero energy is given by rotationally symmetric
cones centered at the Dirac points.  As the hopping energies become different
from each other, those cones are stretched and the Dirac points move, but
still the density of states vanishes linearly close to zero energy ($E\ll
E_{\rm M_{1}},E_{\rm M_{2}}$).  In particular for $t_1=t_2>t_3/2$ the low
energy spectrum is given by $E_{\k_D+\q}=\sqrt{v_x^2 q_x^2+ v_y^2 q_y^2}$,
$v_x=a \sqrt{4 t_1^2-t_3^2}/2$, $v_y=a t_3 \sqrt{3}/2$, resulting in a density
of states $\rho(E)=g_v E/2\pi v_x v_y$, where the degeneracy factor $g_v=2$
takes into account the two Dirac points.  We note that the velocities
$v_x,v_y$ are different if the hoppings are different.  While $v_y$ increases
with increasing $t_3$, $v_x$ decreases and cancels at the phase transition
$t_3=2t_1=2t_2$.  Examples of the density of states in the semimetallic phase
are given in Fig.~\ref{fig:Dos}a.

In the opposite limit $t_3>t_1+t_2$ the energy has its minimum at M$_3$ and
the system is gapped by $2\Delta$ where $\Delta=E_{\rm M_3}=t_3-t_1-t_2$.
Correspondingly the density of states is zero for $E<\Delta$ and has a finite
step at $E=\Delta$.  For $t_1=t_2<t_3/2$ the low energy spectrum is given by
$E_{\k_D+\q}=(\Delta^2+v_x^2 q_x^2 +v_y^2 q_y^2)^{1/2}$, $v_x=a \sqrt{\Delta
  t_1/2}$, $v_y=a \sqrt{3 t_1 t_3/2}$, leading to a density of states
$\rho(E)=\theta(E-\Delta) E/2\pi v_x v_y$, where $\theta(x)$ denotes the
Heaviside step function. This situation is depicted in Fig.~\ref{fig:Dos}c.

Right at $t_3=t_1+t_2$ the two Dirac cones merge at M$_3$ with $E_{\rm M_3}=0$, so
that M$_3$ is the minimum of the conduction band, but there is still no band
gap.  In the particular case $t_1=t_2=t_3/2$, the density of states can be mostly
derived analytically:
\begin{eqnarray}
E_{\rm {\bf M}_3+\q}&=&\frac{t_3}{2} \left[6+2 \cos(q_x a)-8\cos(q_x a/2)\cos(\sqrt{3}
    a q_y/2)\right]^{1/2}\\
\rho(E)&=&\frac{4 \sqrt{\epsilon}}{\pi^2 A_u t_3 }\int_0^1\frac{dx}{\sqrt{1-x^2}\sqrt{1+\epsilon+x^2(1-\epsilon)}\sqrt{1-4\epsilon(1-\epsilon)x^2}}\label{DosPT}
\end{eqnarray}
with $\epsilon=E/2t_3$ denoting the energy in units of the bandwidth.  This
situation is depicted in Fig.~\ref{fig:Dos}b. We note that
$E_{\rm {\bf M}_3+\q}\simeq t_3 [3(a q_y/2)^2+(a/2)^4(q_x^4-6 q_x^2 q_y^2-3q_y^4)/4]^{1/2}$, which shows that the
confinement in the $x$-direction (along which the Dirac points have merged) is much weaker
than in the $y$-direction. Equation~(\ref{DosPT}) is valid for the whole energy range
$0\leq \epsilon \leq 1$. An expansion around $E=0$ results in:
\begin{eqnarray}
\rho(E)&\simeq& \frac{4 \sqrt{\epsilon}}{A_u t_3
  \pi^2}\int_0^1\frac{dx}{\sqrt{1-x^4}}=\sqrt{\epsilon}\frac{4
  \sqrt{\pi}\Gamma(5/4)}{t_3 A_u \pi^2\Gamma(3/4)}\simeq \frac{0.53}{A_u t_3}
\sqrt{\frac{E}{2 t_3}}\label{eq:PT}
\end{eqnarray}
Since the square root has a diverging slope at the origin, it can be viewed as
a transient behavior between the linear density of states of the semimetallic
phase and the step found for the insulating phase. We note that for
$\epsilon=0.5$, which is the energy of the two saddle points $E_{\rm
  M_1}=E_{\rm M_2}$, the integrand of Eq.~(\ref{eq:PT}) contains the factor
$(1-x)^{-1}$ which results in a logarithmic divergence in the density of
states.



\section{Specific heat and structure factor}
Finally we discuss two experimental signatures of the phase transition
expected when fermions populate the lattice at half filling, namely,
 the structure factor accessible by Bragg spectroscopy\cite{Duan07} and the
temperature dependence of the specific heat\cite{Ensher96}.
The specific heat is defined by $c_V\equiv\partial u/\partial
T=\frac{\partial}{\partial T} \int\,dE \rho(E) E f(E)$, where $f(E)$ denotes
the Fermi distribution, $u$ the energy density and $\rho(E)$ the density of
states as calculated in the previous section.  We limit ourselves to the case
of half-filling, where the chemical potential $\mu=0$ for all temperatures
because of the symmetry of the energy around $E=0$ inherent to the employed
tight-binding approximation.

The resulting equations are:
\begin{eqnarray}
  u&=&-\int_0^\infty\,dE \rho(E) E \tanh(E/k_B T)\;;\quad
  c_V=\frac{1}{2k_B T^2}\int_0^\infty\,dE \frac{\rho(E) E^2}{\cosh^2(E/2k_B T)}
\end{eqnarray}
where we have used $\rho(E)=\rho(-E)$. 
This results in a different temperature dependence of the specific heat for
the different phases:
\begin{eqnarray}
  c_V\simeq\left\{ \begin{array}{c c c} 
 T^2\, 9 g_v k_B^3 \zeta(3)/2\pi v_x v_y &;\; \text{for}& \rho(E)=g_v
E/2\pi v_x v_y\\[1.5 ex]
\left(k_B T/t_3\right)^{3/2}\, 2.1 k_B/A_u &;\; \text{for}& \rho(E)\simeq
(E/2 t_3)^{1/2}\, 0.53/A_u t_3 \\[1.5ex]
\exp(-\Delta/k_B T)\,(6T^3+6T^2 \Delta +3 T \Delta^2+\Delta^3)\, 2 k_B^3 /2\pi
v_x v_y T &;\; \text{for}& \rho(E)\simeq \theta(E-\Delta)E/2\pi v_x v_y
\end{array}\right.
\end{eqnarray}
where $\zeta$ denotes the Riemann Zeta Function. Interaction induced
corrections to the specific heat in the semimetallic phase have recently been
studied\cite{Vafek07}.

As pointed out in Ref.~\onlinecite{Duan07} another way to experimentally
visualize the phase transition is to measure the structure factor by Bragg
spectroscopy. We note that the structure factor is up to a constant given by
the imaginary part of the dynamical susceptibility which is well known for
graphene\cite{Guinea94,Wunsch06,Hwang07}. In fact within the semimetallic
phase, the results for the susceptibility of graphene can be easily adapted to include asymmetric
hoppings. Noting that $E_{\q}\simeq \left(v_x^2 q_x^2+v_y^2 q_y^2\right)^{1/2}=\kappa$,
with $\kappa_x=q_x v_x, \kappa_y=q_y
v_y,\kappa=(\kappa_x^2+\kappa_y^2)^{1/2}$, one can reuse all formulae of the
dynamic susceptibility of arbitrary doping by
setting $v_F=1$ and replacing $q\to \kappa$.

As an illustration, we state the result for half-filling:
\begin{eqnarray}
  S(\q,\omega)&=&\frac{g}{16\pi}\frac{v_x^2 q_x^2+v_y^2
    q_y^2}{v_x v_y \sqrt{\omega^2 - v_x^2 q_x^2-v_y^2 q_y^2}}\theta(\omega^2-
  v_x^2 q_x^2-v_y^2 q_y^2)\label{StructSM}
\end{eqnarray}
Here $g=2$ is due to the two Dirac points (valley degeneracy).  For
$t_1=t_2>t_3/2$ the velocities are given by $v_x=a \left(4
  t_1^2-t_3^2\right)^{1/2}/2$, $v_y=a t_3 \sqrt{3}/2$.  We note that in
Ref.~\onlinecite{Duan07} only the energy spectrum was considered while the
overlap factor $f^{\lambda' \lambda}(\k,\q)$ resulting from the spinor
eigenfunctions was ignored. Therefore the $\omega^2$ term present in the
numerator of Eq.~(8) of Ref.~\onlinecite{Duan07} is absent in
Eq.~(\ref{StructSM}) above. We note that $f^{\lambda'
  \lambda}(\k,\q)=|u_{\k+\q,\lambda'}^* u_{\k \lambda}|^2$ describes how easily
the spinor wavefunction $u_{\k \lambda}$ ($\lambda=\mp 1$ labels valence
and conduction band) can be scattered in to $u_{\k+\q,\lambda'}$ and is
responsible for example for the absence of backscattering and the
Klein paradox in graphene.\cite{Katsnelson06} The general form of the spinor
wavefunction for the tight-binding Hamiltonian~(\ref{eq:Hasym}), is
$u_{\k\,\lambda}=2^{-1/2}(-\lambda\, \xi_\k/|\xi_\k|,1)$.

For the gaped case $t_1=t_2<t_3/2$, an analytical formula for the structure
factor at half-filling and for energies close to the gap $E_\q\approx
\left(\Delta^2+v_x^2 q_x^2+v_y^2 q_y^2\right)^{1/2}\simeq \Delta+\frac{v_x^2}{2\Delta}
q_x^2+\frac{v_y^2}{2\Delta} q_y^2$ is given by:
\begin{eqnarray}
S(\q,\omega)&=&\theta(\omega-2\Delta-\frac{v_x^2 q_x^2}{4 \Delta}-\frac{v_y^2 q_y^2}{4 \Delta})\frac{3a^2 q_y^2}{4 (t_3/t_1-2)^2}\frac{\Delta}{2 \pi v_x
  v_y}
\end{eqnarray}
While the step function originates from the single particle eigenspectrum, the
whole $\q$-dependence arises from the spinor character of the wavefunctions
captured in the factor $f^{\lambda' \lambda}(\k,\q)$ described above.

\section{ Topological properties}
Interestingly, a gap only opens after the two Dirac points have merged.
Mathematically it is straightforward to show that an energy minimum of the
conduction band that is not located at one of the symmetry points
M$_1$, M$_2$, M$_3$, $\Gamma$ is automatically a Dirac point.
To show this, we note that a minimum of the conduction band 
is a root of $\nabla_\k E_\k^2=\nabla_\k \left(\eta_\k \eta_\k^*\right)$, with
$\eta_\k=t_3 + t_1 \exp(i\k \a_1)+ t_2 \exp(i\k \a_2)$.
\begin{eqnarray}
  \nabla_\k \left(\eta_\k \eta_\k^*\right)&=&\v_1 {\rm Re}(\eta_\k)+\v_2 {\rm Im}(\eta_\k)\\
  \v_1&=&2 \nabla_\k {\rm Re}(\eta_\k)=-2\left[t_1 \sin(\a_1\cdot \k) \a_1+t_2 \sin(\a_2\cdot \k) \a_2\right]\notag\\
  \v_2&=&2 \nabla_\k {\rm Im}(\eta_\k)=2\left[t_1 \cos(\a_1\cdot \k) \a_1+t_2 \cos(\a_2\cdot \k) \a_2\right]\notag
\end{eqnarray} 
Since $\a_1,\a_2$ are linear independent, $\v_1,\v_2$ are also linear
independent, unless $\sin\left((\a_2-\a_1)\cdot\k)\right)=0$, a condition is
only fulfilled at the symmetry points M$_1$,M$_2$,M$_3$,$\Gamma$.  This
analysis shows that a minimum which is not located at a symmetry point, is
necessarily a
Dirac point, since then both the imaginary and the real parts of $\eta_\k$ vanish and
therefore the energies are $E_{\mp, \k}=\mp
|\eta_\k|=0$. Furthermore, these minima always occur in pairs due to time
reversal symmetry (inversion symmetry around the $\Gamma$-point).

The deeper reason for the stability of the Dirac points is the momentum space
topology of the Bloch wavefunctions\cite{Guinea07,Wu08c}.  This is nicely
illustrated by noting that the Berry phase, defined as\cite{Berry83} $\phi_B=i
\oint_{S} d\k \langle u_\k| \nabla_\k | u_\k\rangle$ with $S$ denoting a
closed path in reciprocal space, takes values $\pm \pi,0$ depending on whether
$S$ encloses one Dirac point, or the other, or both.  Thus each Dirac
point is a source of a $\pm \pi$ delta-function flux of Berry curvature, and
this flux remains invariant under perturbations that preserve space and time inversion
symmetry\cite{Guinea07,Wu08c}. Only if the two Dirac points merge the Berry
curvature vanishes within the whole first Brillouin zone and a gap can open.

We note that the Dirac points are unstable against perturbations that break
spatial invariance. Examples are substrate-induced gap opening in graphene due
to breaking of sublattice symmetry\cite{Zhou07} or honeycomb lattices where
nearest-neighbor hoppings are periodically modified to form a Kekul\'e
pattern\cite{Mudry07}.

\section{ Summary and conclusions}

We have studied the electronic structure and the phase diagram of
non-interacting fermions confined to hexagonal optical lattices. In the first
part of the paper, we have analyzed the appearance of Dirac points in the band
structure of fermionic atoms populating triangular and honeycomb optical
lattices for different confinement strengths.  Numerical results were
complemented by analytical equations both for strong and weak potentials.  The
Dirac points arising in honeycomb and triangular lattices differ in several
important aspects. While in honeycomb lattice the Dirac cone is isolated, it
resonates with a continuum band for a triangular lattice.  Furthermore, in
honeycomb lattices the Fermi-energy coincides with the Dirac point for integer
filling, while a non-integer filling is needed in the case of a triangular
lattice.  For the honeycomb lattice the first pair of Dirac points arises from
the touching of the two lowest bands, which within a tight-binding
description stem from the ground state of the two potential minima per
unit cell.  By contrast, for a triangular lattice the first pair of Dirac
points occurs at the touching of the second and third band, which
within a tight-binding description are formed by the doubly degenerate
first excited state of the single potential minimum per unit cell.

In the second part of this work, we have focused on the phase diagram and the
evolution of the Dirac points in honeycomb lattices adopting a tight-binding
description with asymmetric nearest-neighbor hoppings. The semimetallic phase
is not only realized for strictly symmetric hoppings but rather for an
extended region of the hopping parameter space. This stability of the
semimetallic phase is based on topology. With increasing asymmetry the Dirac
points approach each other along continuous trajectories and at a critical
asymmetry they finally merge and an energy gap opens. We derive analytic
formulae for the trajectories of the Dirac points as well as for the density
of states. Right at the phase transition the density of states increases like
$\sqrt{E}$, which interpolates between the linear dependence of the
semimetallic phase and the step-like increase of the insulating phase.  We
also show how the phase transition becomes manifest in a change of the
temperature dependence of the specific heat as well as in the properties of
the structure factor.

An interesting follow-up of this work concerns the influence of
interactions on the phase diagram, since optical lattices permit a
systematic control of the effective interaction strength by using
Feshbach resonances or modifying the lattice properties\cite{Bloch08}.

\section*{Acknowledgments}
We appreciate helpful discussions with C. Creffield, A. Cortijo, L. Brey and
L. Amico.  This work has been supported by the EU Marie Curie RTN Programme
No.  MRTN-CT-2003-504574, the EU Contract 12881 (NEST), by MEC (Spain) through
Grants No. FIS2004-05120, FIS2005-05478-C02-01, FIS2007-65723, and by the Comunidad de Madrid
through program CITECNOMIK.

\appendix
\section{Dirac cones in weak symmetric lattices}\label{app:NFE}
Here we apply the nearly-free-particle approximation to derive the
eigenspectrum determined by Eq.~(\ref{Eq:Eigenspectrum}) with $V_1=V_2=V_3=V$
close to the K-point.  In absence of any potential the dispersion relation
at the K-point is three-fold degenerate, where the three states can be
labeled by the $\k$-values $\k\in\{\K,\K-\G_{10},\K-\G_{11}\}$.  Treating the
lattice within lowest order perturbation theory, the eigenvalue problem at
$\k=\K+\q$ has then the following form:
\begin{eqnarray}
\left(\begin{array}{c c c} E^0_{\K+\q} & V/2 & V/2 \\ V/2 &
    E^0_{\K-\G_{10}+\q}& V/2 \\ V/2& V/2 & E^0_{\K-\G_{11}+\q}
  \end{array}\right)\left(\begin{array}{c}c_{\K+\q}\\ c_{\K-\G_{10}+\q}\\
    c_{\K-\G_{11}+\q}\end{array}\right)=E_{\K+\q}
\left(\begin{array}{c}c_{\K+\q}\\ c_{\K-\G_{10}+\q}\\ c_{\K-\G_{11}+\q}\end{array}\right)
\end{eqnarray}
The solutions of the above eigenproblem for $\q=0$ are given by $E=E^0_\K+V$
with eigenvector $3^{-1/2}(1,1,1)$ and $E=E^0_\K-V/2$, which is two fold
degenerate with a possible set of orthonormal eigenvectors given by:
$\v_1=2^{-1/2}(-1,1,0)$ and $\v_2=6^{-1/2}(-1,-1,2)$.  We note that the
remaining two-fold degeneracy is protected by topology and thus persists
beyond a perturbative description. This degeneracy occurs in the ground state
for $V>0$ (honeycomb lattice) and in the first excited state for $V<0$
(triangular lattice).

The remaining two-fold degenaracy of $\v_1,\v_2$ disappears for finite $\q$.
The effective two-band Hamiltonian spanned by $\v_1$,$\v_2$ is given by:
\begin{eqnarray}
\left(H_{\rm eff}(\q)\right)_{i j}&=&\langle v_i|H(\q)|v_j\rangle\\
H_{\rm eff}(\q)&=&\left(E^0_\K-V/2+E^0_\q\right) \mathbbm{1}+\hbar v_F \left(-\frac{q_x}{2}+\frac{\sqrt{3}}{2}
q_y\right)\sigma_z + \hbar v_F \left(\frac{\sqrt{3}}{2} q_x+\frac{1}{2} q_y\right)\sigma_x
\end{eqnarray}
with $v_F=2\pi \hbar/3a m$. The eigenenergies are given by $E_{\K+\q}=E^0 \pm
\hbar v_F q $ with $E_0=E^0_{\K}-V/2+E^0_\q$. We note that the effective
Hamiltonian can be transformed by a unitary transformation to $\tilde{H}_{\rm
  eff}=E_0\mathbbm{1} + \hbar v_F (q_x \sigma_x+ q_y \sigma_y)$, which is
identical (except for the value of $v_F$) to the low-energy Hamiltonian
obtained in the tight-binding approximation .

\section{Tight-binding description for symmetric potentials}\label{app:TB}
\subsection{Symmetric honeycomb lattice}\label{app:DOSCrossover}
For the honeycomb lattice there are two degenerate minima per unit cell and
for sufficiently deep potential wells [i.e. for $V\ll \frac{\hbar^2}{2 m
  a^2}$, with $m$ the atomic mass and $V$ the confinement amplitudes of
Eq.~(\ref{eq:Pot})] the basis for the two lowest bands can be restricted to
the ground state of each minimum. In the symmetric case (equal laser
intensities) the potential around each minimum is rotationally symmetric and
harmonic and the minima form a honeycomb lattice. Thus the nearest-neighbor
hoppings $t$ are all the same.
\begin{eqnarray}
H&=&-t \sum_{i} a^\dag(\r_i)\sum_j b(\r_i+\d_j) + {\rm H.c.} =-\sum_\k \left(a_\k^\dag,b_\k^\dag\right)\left(\begin{array}{c c}0 & \xi_\k\\
  \xi_\k^* & 0\end{array}\right) \left(\begin{array}{c} a_\k\\
  b_\k\end{array}\right)\,;\quad \xi_\k=t \sum_j \exp(-i \k\cdot\d_j) \label{eq:Hsym}
\end{eqnarray}
Here $a(\r)$ destroys a particle at the A-site at position $\r$ and the $\d_j$
denotes the vectors connecting an A-site with its neighboring B-sites, see
Fig.~\ref{fig:Unitcells}b. Fourier transformation is given by $a(\r_i)=N^{-1/2}
\sum_\k \exp(i\k\cdot\r_i) a_\k$, where $N$ is the number of unit cells.  The
eigenenergies of the above Hamiltonian are given by: $E_\k=\pm |\xi_k|=\pm t
[3+2\cos(k_y a)+4\cos(\sqrt{3} a k_x/2)\cos(a k_y/2)]^{1/2}$.  The energy
spectrum is symmetric around $E=0$, where the Fermi surface only consists of
singular (Dirac) points located at the K-points.  Close to the K-point one
can approximate $\xi_{\K+\q}\simeq -v_F(q_x - i q_y)$ with $v_F=\sqrt{3} a
t/2$, so that the Hamiltonian close to the K-point can be written as
$H_{\rm eff}=\hbar v_F (q_x \sigma_x+ q_y \sigma_y)$ with eigenenergies $E_\q=\pm
\hbar v_F q$.

\subsection{Symmetric triangular lattice}
The triangular lattice has one minimum per unit cell. For a symmetric lattice
the potential around the minimum is rotationally symmetric and harmonic.  The
ground state of such a minimum is non-degenerate while the first excited state
is two fold degenerate and can be represented by $p_x(x,y)=\psi_1(x)\psi_0(y)$
and $p_y(x,y)=\psi_0(x)\psi_1(y)$, where $x,y$ are measured with respect of
the center of the minimum and $\psi_0(x),\psi_1(x)$ denote the ground state
and the first excited state of the one-dimensional harmonic oscillator.

\begin{figure}
  \begin{center}
    \includegraphics*[angle=0, width=0.5\linewidth]{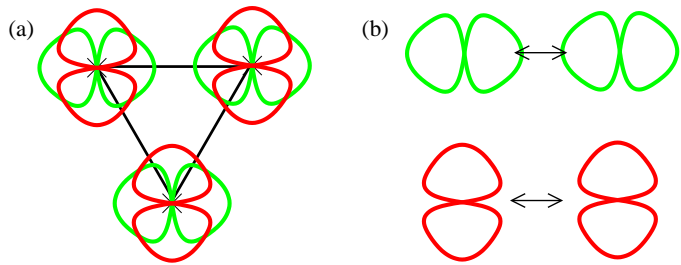} 
    \caption{(a) Triangular lattice with schematic representation of the two
      degenerate excited $\pi$ orbitals. (b) Top $t_\sigma$, bottom $t_\pi$.}
\label{fig:TBTriangle}
\end{center}
\end{figure}

The band arising from the ground state is described by the following tight-binding Hamiltonian:
\begin{eqnarray}
H&=&-t \sum_{\langle i,j\rangle } c^\dag_i c_j=\sum_{\k} E_\k c_\k^\dag
c_\k\,;\quad E_\k=-2t \left[\cos(\k\cdot\a_1)+
  \cos(\k\cdot\a_2)+\cos(\k\cdot(\a_1-\a_2))\right]
\end{eqnarray}

The hopping between the $p_x$ and $p_y$ orbitals is no longer
isotropic\cite{Wu07a}.  As visualized in Fig.~\ref{fig:TBTriangle}b, we
introduce the hopping amplitudes $t_\sigma$ and $t_\pi$ depending on the
relative orientation of the orbitals. Non-parallel hoppings (e.g. from $p_x$
to $p_y$) are excluded by the symmetry of the wavefunctions. 

The tight-binding Hamiltonian describing the two bands arising from the two-fold degenerate
first excited state is then given by:
\begin{eqnarray}
H&=&\sum_{\k}\left(c_\k^\dag,d_\k^\dag\right)  H_k \left(\begin{array}{c}
    c_\k\\d_\k\end{array}\right)\,;\quad H_\k= h_1 \mathbbm{1} + h_2
\sigma_x+h_3\sigma_z\,;\quad E_\k=h_1\pm\sqrt{h_2^2+h_3^2}\\
\left(\begin{array}{c}t_c\\t_r \end{array}\right)
&=&\left(\begin{array}{c c}1/2&1/2\\-1&1
  \end{array}\right)\left(\begin{array}{c}t_\sigma\\t_\pi
  \end{array}\right)\;;\quad h_1=2 t_c \left[\cos(a k_x)+2\cos(a k_x/2)\cos(a k_y\sqrt{3}/2)\right]\notag\\
h_2&=&\sqrt{3} t_r \sin(a k_x/2)\sin(a k_y\sqrt{3}/2)\;;\quad
h_3=t_r\left[\cos(a k_x/2)\cos(a k_y
    \sqrt{3}/2)-\cos(a k_x)\right]\notag
\end{eqnarray}  

We note that both bands touch at the $\Gamma$ point as well as at the K-points
(see Figs~\ref{fig:Unitcells}c and~\ref{fig:Band}).  The energies are given
by $E_\Gamma=6t_c$ and $E_\K=-3 t_c$, while at the M-point the energies are
split with $E_{\rm M}=-2t_c\pm 2t_r$.  Around the K-points $\k=\K+\q$ the energy
dispersion to linear order in $\q$ is rotationally symmetric,
$E_{\K+\q}=-3t_c\pm 3^{3/2} a t_r q/4 +{\cal O}(\q^2)$, so that a circular Dirac cone
is formed around each K-point.  We note that the electronic structure of a
triangular lattice with degenerate $d$-orbitals shows similar
features\cite{Guillamon08}.  
\bibliography{NJP_BWunsch}
\end{document}